\documentclass{svmult}

\usepackage{mathptmx}       
\usepackage{helvet}         
\usepackage{courier}        
\usepackage{type1cm}        
\usepackage{makeidx}         
\usepackage{graphicx}        
\usepackage{multicol}        
\usepackage[bottom]{footmisc}

\usepackage{tabularx}
\usepackage{rotating}

\makeindex             

\begin{document}

\title*{Generative Network Automata: A Generalized Framework for
  Modeling Adaptive Network Dynamics Using Graph Rewritings}
\titlerunning{Generative Network Automata}
\author{Hiroki Sayama and Craig Laramee}

\institute{Hiroki Sayama and Craig Laramee \at Collective Dynamics of
  Complex Systems Research Group / Department of Bioengineering,
  Binghamton University, State University of New York, P.O. Box 6000,
  Binghamton, NY 13902-6000, USA, \email{sayama@binghamton.edu,
    claramee@binghamton.edu} \at Hiroki Sayama is also an Affiliate of
  the New England Complex Systems Institute, 24 Mt. Auburn St.,
  Cambridge, MA 02138, USA, \email{sayama@necsi.edu}}
%
%
\maketitle

\abstract*{A variety of modeling frameworks have been proposed and
  utilized in complex systems studies, including dynamical systems
  models that describe state transitions on a system of fixed
  topology, and self-organizing network models that describe
  topological transformations of a network with little attention paid
  to dynamical state changes. Earlier network models typically assumed
  that topological transformations are caused by exogenous factors,
  such as preferential attachment of new nodes and stochastic or
  targeted removal of existing nodes. However, many real-world complex
  systems exhibit both of these two dynamics simultaneously, and they
  evolve largely autonomously based on the system's own states and
  topologies. Here we show that, by using the concept of graph
  rewriting, both state transitions and autonomous topology
  transformations of complex systems can be seamlessly integrated and
  represented in a unified computational framework. We call this novel
  modeling framework ``Generative Network Automata (GNA)''. In this
  chapter, we introduce basic concepts of GNA, its working definition,
  its generality to represent other dynamical systems models, and some
  of our latest results of extensive computational experiments that
  exhaustively swept over possible rewriting rules of simple
  binary-state GNA. The results revealed several distinct types of the
  GNA dynamics.}

\abstract{A variety of modeling frameworks have been proposed and
  utilized in complex systems studies, including dynamical systems
  models that describe state transitions on a system of fixed
  topology, and self-organizing network models that describe
  topological transformations of a network with little attention paid
  to dynamical state changes. Earlier network models typically assumed
  that topological transformations are caused by exogenous factors,
  such as preferential attachment of new nodes and stochastic or
  targeted removal of existing nodes. However, many real-world complex
  systems exhibit both of these two dynamics simultaneously, and they
  evolve largely autonomously based on the system's own states and
  topologies. Here we show that, by using the concept of graph
  rewriting, both state transitions and autonomous topology
  transformations of complex systems can be seamlessly integrated and
  represented in a unified computational framework. We call this novel
  modeling framework ``Generative Network Automata (GNA)''. In this
  chapter, we introduce basic concepts of GNA, its working definition,
  its generality to represent other dynamical systems models, and some
  of our latest results of extensive computational experiments that
  exhaustively swept over possible rewriting rules of simple
  binary-state GNA. The results revealed several distinct types of the
  GNA dynamics.}

\section{Introduction}

A variety of modeling frameworks have been proposed and utilized for
research on the dynamics of complex systems
\cite{bar-yam97,wiggins03,boccara04}. A major class of modeling
frameworks is that of dynamical systems models, including ordinary or
partial differential equations and iterative maps \cite{strogatz94},
artificial neural networks \cite{mcculloch43,hopfield82}, random
Boolean networks \cite{kauffman69,derrida86,kauffman93}, and cellular
automata \cite{wolfram84,ilachinski01}. While they are capable of
producing strikingly complex and even biological-like behaviors
\cite{may76,berlekamp82,pearson93,sayama99,salzberg04b}, these tools
generally assume a network made of a fixed number of components
organized in a fixed topology. Their dynamics are considered as
trajectories of system states in a confined phase space with
time-invariant dimensions.

The recent surge of network theory in statistical physics has
demonstrated yet another graph-theoretic approach to complex systems
modeling \cite{watts98,strogatz01,newman06}. It addresses the
self-organization of network structure via local topological
transformations such as random or preferential addition, modification
and removal of components and their interactions (i.e., nodes and
links). Among the most actively investigated issues in this field is
how statistical properties of the entire network topology will be
affected by additions (growth or augmentation) and removals (failures
or attacks) of nodes and links, and in particular, how networks can be
more robust against the latter
\cite{albert00,albert02,shargel03,dafontouracosta04,beygelzimer05}. Those
additions and removals are typically assumed as perturbations coming
from external sources, not incorporated into the dynamics of the
network itself. They are also limited in that not much attention has
been paid to dynamical state changes on the network. Researchers
recently started investigating dynamical state changes on complex
networks
\cite{bar-yam04,motter04,deaguiar05,zhou05,tomassini06,motter06}. They
are still largely focusing on fixed network topologies or topologies
varied by exogenous perturbations.

When looking into real-world complex networks, however, one can find
many instances of networks whose states and topologies ``coevolve'',
i.e., they keep changing over the same time scales due to the system's
own dynamics (Table \ref{examples}). In these networks, state
transitions of each component and topological transformations of
networks are deeply coupled with each other. Understanding and
describing the coevolution of states and topologies of networks is now
recognized as one of the most important problems to address
\cite{albert02,gross08}. Several theoretical models of coevolutionary
networks have been proposed and studied most recently
\cite{holme06a,holme06b,gross06,pacheco06,palla07}, yet each of these
studies used different model formulations for different phenomena,
with limited implications given for how these coevolutionary network
models could be linked to other existing complex systems models.

\begin{table}[tp]
\caption{Real-world examples of complex networks whose states and
  topologies change over the same time scales due to the network's own
  dynamics.}
\label{examples}
\newcolumntype{L}{>{\raggedright\arraybackslash}X}

\begin{tabularx}{\columnwidth}{>{\raggedright\arraybackslash}p{18mm}|LLLLL}

\hline \hline

Network & Nodes & Links & Example of node states & Example of node
addition or removal & Example of topological changes \\

\hline
\hline

Organism & Cells & Cell adhesions, intercellular communications &
Gene/protein activities & Cell division, cell death & Cell migration
\\

\hline

Ecological community & Species & Ecological relationships (predation,
symbiosis, etc.) & Population, intraspecific diversities & Speciation,
invasion, extinction & Changes in ecological relationships via
adaptation \\

\hline

Epidemio\-logical network & Individuals & Physical contacts &
Pathologic states & Death, quarantine & Reduction of physical contacts
\\

\hline

Social network & Individuals & Social relationships, conversations,
collaborations & Sociocultural states, political opinions, wealth &
Entry to or withdrawal from community & Establishment or renouncement
of relationships \\

\hline

\end{tabularx}
\end{table}

Here we aim to address the above-mentioned lack of linkages between
coevolutionary network models and other existing complex systems
models by developing a more comprehensive formulation. Specifically,
we show that, by using the concept of graph rewriting, both state
transitions and autonomous topology transformations of complex systems
can be seamlessly integrated and represented in a unified
computational framework. We call this novel modeling framework ``{\em
  Generative Network Automata (GNA)}'' \cite{sayama07}. The name
indicates the integration of knowledge accumulated in dynamical
systems theory, network theory, and graph grammar theory.

In the following sections, we will introduce basic concept of graph
rewriting, a working definition of GNA, its generality to represent
other dynamical systems models, and some of our latest results of
extensive computational experiments that exhaustively swept over
possible rewriting rules of simple binary-state GNA. The results
revealed several distinct types of the GNA dynamics.

\section{About Graph Rewriting}

The key characteristic of GNA is that it should have mechanisms for
transformations of local network topologies as well as transitions of
local states. Topological transformations may be modeled as a
rewriting process of local network configurations. We will therefore
adopt methods and techniques developed in graph grammar theory
\cite{rozenberg97} to construct general formulations of GNA.

Graph grammars, studied since late 1960's in theoretical computer
science \cite{ggproc1,ggproc2,ggproc3,ggproc4}, are an extension of
formal generative grammars in computational linguistics to discuss
similar rule-based generative processes of graphs, or networks. They
recursively define a set of ``valid'' graph topologies that can be
generated through repetitive applications of a given set of node
and/or link replacement rules. A computational implementation of such
processes is called a graph rewriting system, often used to simulate
particular generative processes of network topology. Here the word
``generative'' means that the replacements are triggered by local
topological features of the network itself, and not by external
sources of perturbation as typically assumed in modern network theory.

A classic, and probably most widely known, example of graph rewriting
systems is the Lindenmayer system, or L-system
\cite{lindenmayer68}. It is a simple rewriting system that can produce
self-similar recursive structures in a sequential string (in this
sense, the L-system remains within the range of classic formal
grammars). What makes this system outstanding is that it comes with an
interpretation that converts a resultant string into a tree-like
topological structure, which may appear just like a natural tree if
parameters are appropriately chosen. This example shows the capability
of graph rewriting systems to describe the emergence of natural
complex structures using a set of small local rules.

Although their relevance to biology was initially recognized
\cite{ggproc1,doi84}, applications of graph grammars have so far
remained within computer science, such as pattern recognition,
compiler design, and data type and process specification
\cite{rozenberg97,ggproc2,ggproc3,ggproc4}, and their use has been not
so common even within computer science due to unintuitive, complicated
formulation and lack of software tools for modeling
\cite{blostein96}. Moreover, most applications were primarily focused
on context-free rewriting rules, and they rarely considered dynamical
state transitions on networks. Recently, context-dependent graph
grammars have been applied to describe reaction rules in artificial
life/artificial chemistry, including models of self-replication
\cite{tomita02,hutton02,klavins04a}, self-assembly \cite{klavins04b},
morphogenesis \cite{kniemeyer04,kurth05} and dynamic state changes
\cite{kurth05} of artifacts. However, none of them integrated graph
grammars into complex systems modeling in a flexible, generalizable
way so as to be readily applicable to networks studied in other
domains.

To the best of our knowledge, our GNA framework is among the first to
systematically integrate graph rewritings in the representation and
computation of the dynamics of complex networks that involve both
state transition and autonomous topological transformation. Our
long-term goal is to develop a comprehensive theory of GNA and a set
of analytical/computational tools that can be broadly applied to the
modeling of various complex systems.

\section{Definition of GNA}

A working definition of GNA is a network made of dynamical nodes and
directed links between them. Undirected links can also be represented
by a pair of directed links symmetrically placed between nodes. Each
node takes one of the (finitely or infinitely many) possible states
defined by a node state set $S$. The links describe referential
relationships between the nodes, specifying how the nodes affect each
other in state transition and topological transformation. Each link
may also take one of the possible states in a link state set $S'$. A
configuration of GNA at a specific time $t$ is a combination of states
and topologies of the network, which is formally given by the
following:
\begin{itemize}
\item $V_t$: A finite set of nodes of the network at time $t$. While
  usually assumed as time-invariant in conventional dynamical systems
  theory, this set can dynamically change in the GNA framework due to
  additions and removals of nodes.
\item $C_t:V_t \to S$: A map from the node set to the node state set
  $S$. This describes the global state assignment on the network at
  time $t$. If local states are scalar numbers, this can be
  represented as a simple vector with its size potentially varying
  over time.
\item $L_t:V_t \to \{V_t \times S'\}^*$: A map from the node set to a
  list of destinations of outgoing links and the states of these
  links, where $S'$ is a link state set. This represents the global
  topology of the network at time $t$, which is also potentially
  varying over time.
\end{itemize}

States and topologies of GNA are updated through repetitive graph
rewriting events, each of which consists of the following three steps:
\begin{enumerate}
\item Extraction of part of the GNA (subGNA) that will be subject to
change.
\item Production of a new subGNA that will replace the subGNA selected
above.
\item Embedding of the new subGNA into the rest of the whole GNA.
\end{enumerate}
The temporal dynamics of GNA can therefore be formally defined by the
following triplet $\langle E,R,I \rangle$:
\begin{itemize}
\item $E$: An extraction mechanism that determines which part of the
  GNA is selected for the updating. It is defined as a function that
  takes the whole GNA configuration and returns a specific subGNA in
  it to be replaced. It may be deterministic or stochastic.
\item $R$: A replacement mechanism that produces a new subGNA from the
  subGNA selected by $E$ and also specifies the correspondence of
  nodes between the old and new subGNAs. It is defined as a function
  that takes a subGNA configuration and returns a pair of a new subGNA
  configuration and a mapping between nodes in the old subGNA and
  nodes in the new subGNA. It may be deterministic or stochastic.
\item $I$: An initial configuration of GNA.
\end{itemize}

\begin{figure}[tp]
\centering
\includegraphics[width=0.6\columnwidth]{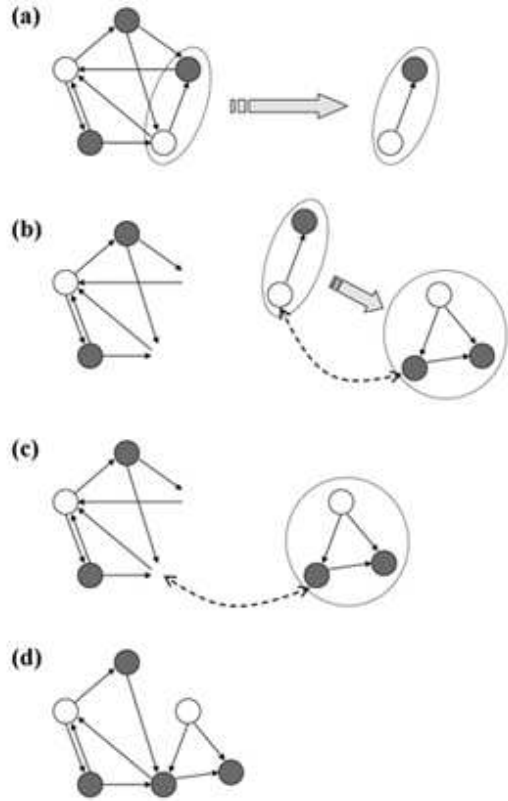}
\caption{GNA rewriting process. (a) The extraction mechanism $E$
selects part of the GNA. (b) The replacement mechanism $R$ produces a
new subGNA as a replacement of the old subGNA and also specifies the
correspondence of nodes between old and new subGNAs (dashed
line). This process may involve both state transition of nodes and
transformation of local topologies. The ``bridge'' links that used to
exist between the old subGNA and the rest of the GNA remain
unconnected and open. (c) The new subGNA produced by $R$ is embedded
into the rest of the GNA according to the node correspondence also
specified by $R$. In this particular example, the top gray node in the
old subGNA has no corresponding node in the new subGNA, so the bridge
links that were connected to that node will be removed. (d) The
updated configuration after this rewriting event.}
\label{rewriting}
\end{figure}

There are a couple of other commonly used procedures needed to
simulate GNA dynamics, such as the removal of the selected subGNA from
the whole GNA and the re-connection of ``bridge'' links (i.e., links
that were between the old subGNA and the rest of the GNA) when
embedding the new subGNA. Because the workings of these procedures are
fairly obvious, we omit detailed explanations for them. The above $E,
R, I$ are sufficient to uniquely define specific GNA models. The
entire picture of a rewriting event is illustrated in
Fig.~\ref{rewriting}, which visually shows how these mechanisms work
together.

This rewriting process, in general, may not be applied synchronously
to all nodes or subGNAs in a network, because simultaneous
modifications of local network topologies at more than one places may
cause conflicting results that are inconsistent with each other. This
limitation will not apply though when there is no possibility of
topological conflicts, e.g., when the rewriting rules are all
context-free, or when GNA is used to simulate conventional dynamical
networks that involve no topological changes.

We note that it is a unique feature of GNA that the mechanism of
subgraph extraction is explicitly described in the formalism as an
algorithm $E$, not implicitly assumed outside the grammatical rules
like what other graph rewriting systems typically adopt (e.g.
\cite{kurth05}). Such algorithmic specification allows more
flexibility in representing diverse network evolution and less
computational complexity in implementing their simulations,
significantly broadening the areas of application. For example, the
preferential attachment mechanism widely used in modern network theory
to construct scale-free networks is hard to describe with pure graph
grammars but can be easily written in algorithmic form in GNA, as
demonstrated in the next section.

While the definition given above is one of the simplest possible
formulations of GNA, it already has considerable complexity compared
to conventional dynamical systems models. The possibility of temporal
changes of $V_t$ and $L_t$ particularly makes it difficult to
investigate its dynamical properties analytically. However, the
updating process of GNA is algorithmically described and hence their
dynamics can be experimented through computer simulation relatively
easily. We have developed a package in Wolfram Research Mathematica for
small-scale simulation and visualization of GNA with node
states\footnote{The Mathematica package is still under active
  development but may be available upon request.}. The results
presented in this chapter were obtained using this package.

\section{Generality of GNA}

The GNA framework is highly general and flexible so that many existing
dynamical network models can be represented and simulated within this
framework.

\begin{figure}[tp]
\centering \includegraphics[width=0.95\columnwidth]{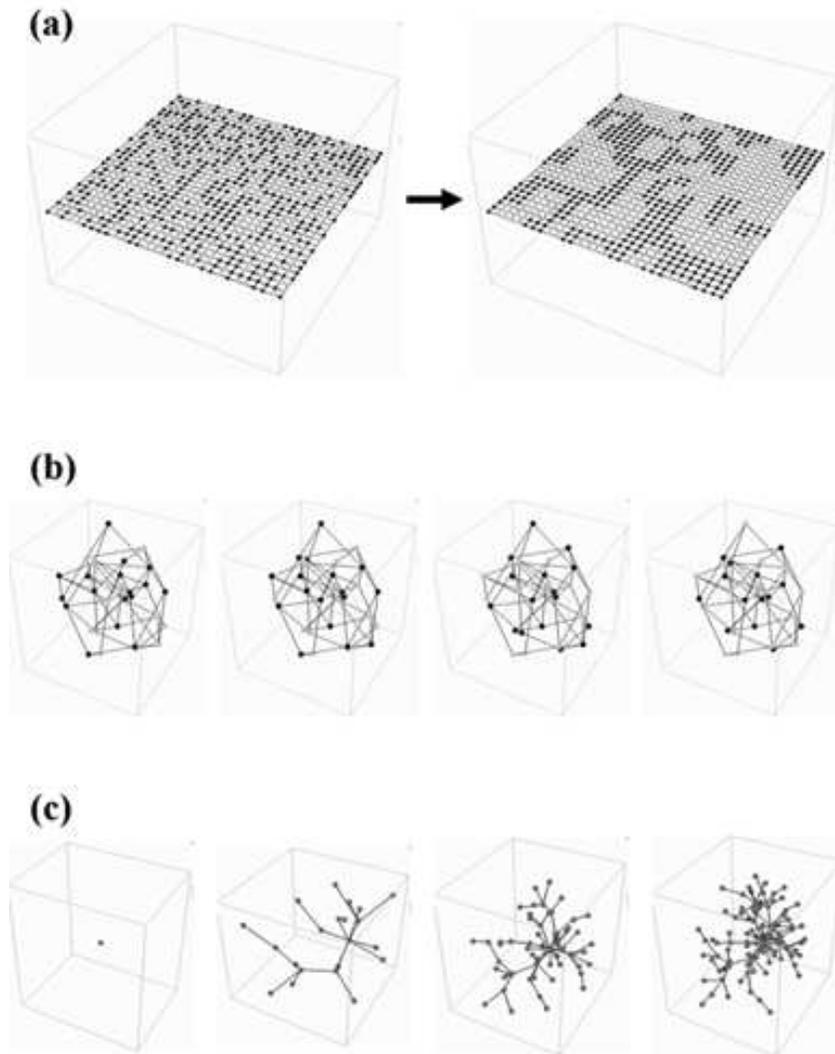}
\caption{Various dynamical network models simulated using GNA. These
  examples were represented in the same format of $\langle E, R, I
  \rangle$ (see text) and simulated using the same simulator package
  implemented in Mathematica. (a) Simulation of asynchronous 2-D
  binary cellular automata with von Neumann neighborhoods and local
  majority rules. Space size: $100 \times 100$. (b) Simulation of an
  asynchronous random Boolean network with $N=30$ and $K=2$. Time
  flows from left to right. Nodes of random Boolean networks are
  non-homogeneous, i.e., they obey different state-transition
  rules. Here each node's own state-transition rule is embedded as
  part of its state, and the replacement mechanism $R$ refers to that
  information when calculating the next state of a node. (c)
  Simulation of a network growth model with the Barab\'{a}si-Albert
  preferential attachment scheme. Time flows from left to right. Each
  new node is attached to the network with one link. The extraction
  mechanism $E$ is implemented so that it determines the place of
  attachment preferentially based on the node degrees, which causes
  the formation of a scale-free network in the long run.}
\label{applications}
\end{figure}

For example, if $R$ always conserves local network topologies and
modifies states of nodes only, then the resulting GNA is a
conventional dynamical network model, including cellular automata,
artificial neural networks, and random Boolean networks
(Fig.~\ref{applications} (a), (b)). A straightforward application of
GNA typically comes with asynchronous updating schemes, as introduced
in the previous section. Since asynchronous automata networks can
emulate any synchronous automata networks \cite{nehaniv04}, the GNA
framework covers the whole class of dynamics that can be produced by
conventional dynamical network models. Moreover, as mentioned earlier,
synchronous updating schemes could also be implemented in GNA for this
particular class of models because they involve no topological
transformation.

On the other hand, many network growth models developed in modern
network theory can also be represented as GNA if appropriate
assumptions are implemented in the subGNA extraction mechanism $E$ and
if the replacement mechanism $R$ causes no change in local states of
nodes (Fig.~\ref{applications} (c)).

\section{Computational Exploration of Possible Dynamics of Simple
Binary-State GNA}

In this section we report our latest results of extensive
computational experiments that exhaustively swept over possible
rewriting rules of simple binary-state GNA. The results shown here
were obtained with much less restricted rule sets than those assumed
in our previous work \cite{sayama07}.

\subsection{Assumptions}

\begin{figure}[tp]
\centering
\includegraphics[scale=0.165]{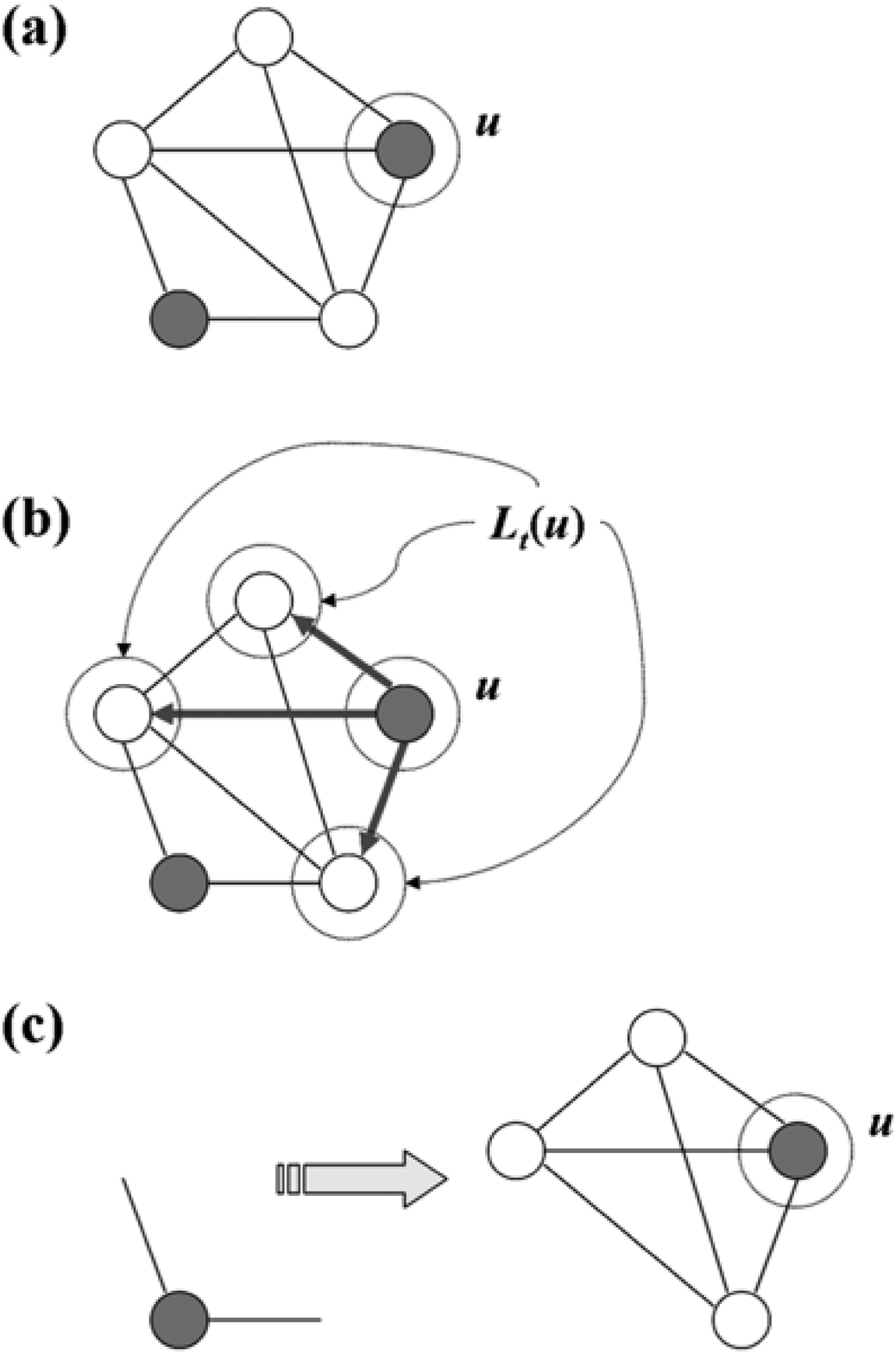}
~~~~~
\includegraphics[scale=0.165]{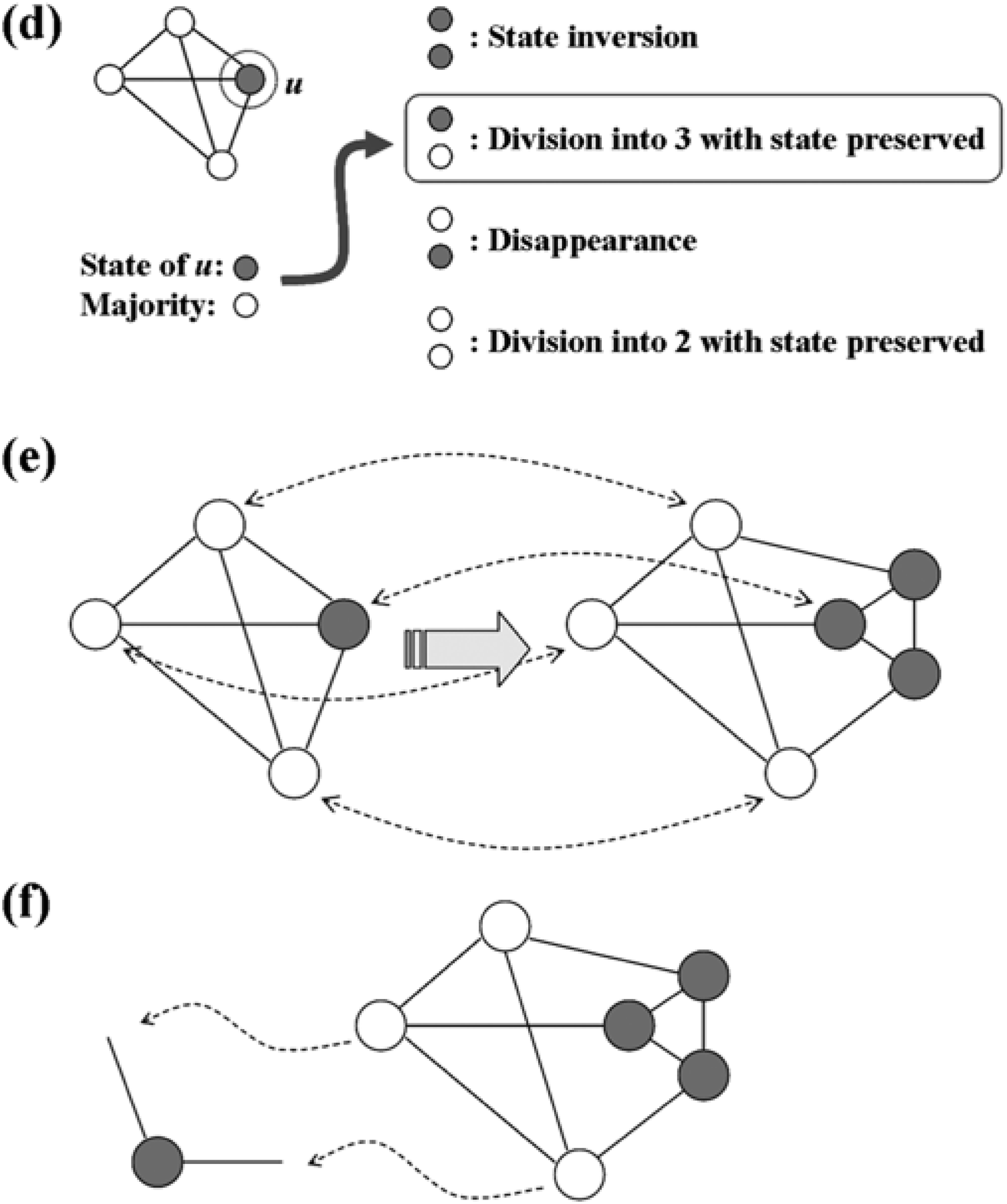}
\caption{Simplified GNA rewriting process used for the exhaustive
sweep experiments. The extraction mechanism $E$ (a) randomly picks one
node $u$, (b) takes all the destination nodes of its outgoing links
$L_t(u)$, and (c) produces a subGNA induced by those nodes $\{u\} \cup
L_t(u)$. The replacement mechanism $R$ (d) refers to the state of the
central node $u$ in the selected subGNA and the local majority state
within it to determine what happens to the local configuration, (e)
produces a new subGNA as well as the correspondence of nodes between
the old and new subGNAs based on the choice made in (d), and then (f)
embeds the new subGNA into the rest of the GNA.}
\label{simplerewriting}
\end{figure}

There are infinitely many possible mechanisms for $E$ and $R$ because
there are no theoretical upper bounds in terms of the size of the old
subGNA selected by $E$ (it could be infinitely large as the GNA grows)
and the new subGNA produced by $R$ (it could be arbitrarily large by
the design of $R$). Making reasonable assumptions to restrict the
possibility of mechanisms for $E$ and $R$ is critical to facilitate
systematic study on the dynamics of GNA. Here we make the following
assumptions (Fig.~\ref{simplerewriting}):
\begin{enumerate}
\item Node states are binary (0 or 1).
\item No link state is considered (i.e., links homogeneously take only
  one state and it will never change).
\item Links are undirected (i.e., every connection between nodes is
  represented by a pair of symmetrically placed directed links).
\item The extraction mechanism $E$ always selects a subGNA by
\begin{enumerate}
\item randomly picking one node $u$ from the entire GNA
  (Fig.~\ref{simplerewriting} (a)),
\item taking all the destination nodes of its outgoing links $L_t(u)$
  (Fig.~\ref{simplerewriting} (b)), and
\item producing a subGNA ``induced'' by these nodes $\{u\} \cup
  L_t(u)$, i.e., a subGNA that includes all these nodes as well as all
  the links present between them (Fig.~\ref{simplerewriting} (c)).
\end{enumerate}
\item The replacement mechanism $R$ only refers to the state of the
  central node $u$ and the local majority state within the induced
  subGNA. If there are equal numbers of 0's and 1's within the subGNA,
  one of the two states is randomly chosen. This two-bit information
  will be used to determine what will happen to the local
  configuration (Fig.~\ref{simplerewriting} (d)). The following ten
  possible rewriting outcomes are made available (which are extended
  from \cite{sayama07}):
\begin{itemize}
\item [0) ] The central node $u$ disappears.
\item [1) ] Everything remains in the same condition.
\item [2) ] The state of the central node $u$ is inverted.
\item [3) ] The central node $u$ divides into two with the state
  preserved in both nodes.
\item [4) ] The central node $u$ divides into two with the state
  inverted in both nodes.
\item [5) ] The central node $u$ divides into two with the state
  inverted in one node.
\item [6) ] The central node $u$ divides into three with the state
  preserved in all three nodes.
\item [7) ] The central node $u$ divides into three with the state
  inverted in all of three nodes.
\item [8) ] The central node $u$ divides into three with the state
  inverted in two of three nodes.
\item [9) ] The central node $u$ divides into three with the state
  inverted in one of three nodes.
\end{itemize}
In cases where node division occurs, the links that were connected to
the central node $u$ is distributed as evenly as possible to its
daughter nodes (Fig.~\ref{simplerewriting} (e)).
\item The initial condition $I$ consists of a single node with state
  0.
\end{enumerate}
Note that the above model assumptions will always generate planar
graphs in which the node degrees are bounded up to three when
initiated with a single node. Therefore all the results shown in this
chapter are topologically planar.

\subsection{Methods}

We carried out an exhaustive sweep of all the possible rewriting rules
that satisfy the assumptions discussed above. Since the extraction
mechanism $E$ is uniquely defined, it is only the replacement
mechanism $R$ that can be varied. Here $R$ is defined as a function
that maps each of the four possible two-bit inputs to one of the ten
possible actions. Therefore the number of all the possible $R$'s is
$10^{2^2}=10000$. To indicate a specific $R$, we will use its ``rule
number'' $rn(R)$ that is defined by
\begin{equation}
rn(R) = a_{11}\times 10^3 + a_{10}\times 10^2 + a_{01}\times 10^1 +
a_{00}\times 10^0 ,
\end{equation}
where $a_{ij}$ is a numerical representation (numbers associated with
each of the ten possible actions shown above) of the choice that $R$
will make when the state of the central node $u$ is $i$ and the local
majority state is $j$.

It should be noted that there are two different ways of counting time
steps in asynchronous simulations. One is simply to count one
rewriting event as one time step, which we call {\em computational
  time}. The other is to measure the progress of virtual time in a
simulated world between discrete events by considering one rewriting
event as taking $1/N_t$ of the unit of time, where $N_t$ is the number
of nodes at time $t$. This is based on the assumption that every node
is updated once on average per unit of time, which is a reasonable and
useful assumption especially when one wants to compare results of
asynchronous simulations with those of synchronous ones. We call the
latter notion of time {\em simulated time}. All the $t$'s in this
chapter represent simulated time.

We simulated the GNA dynamics for $rn$ ranging from 0 to 9999. For
each $rn$ five independent simulation runs were conducted and the
average of their results were used. Each run continued until 500
rewriting events were simulated, or $N_t$ exceeded 1000, or $N_t$
became 0, whichever was sooner.

During each run, we recorded time series of $N_t$ by sampling its
value in every half unit of simulated time. We then calculated its
growth characteristics, estimated order of polynomial growth $k$ and
estimated rate of exponential growth $r$, by conducting nonlinear
fitting of a hypothetical growth model to the time series data
(explained later). In addition, after each simulation run, we measured
the following quantities of the final GNA configuration:
\begin{itemize}
\item Number of nodes
\item Number of links
\item Average node degree
\item Number of connected components
\item Size of the largest connected component
\item Average node state
\end{itemize}
If all the nodes disappear during the simulation run, the average node
degree and the average node state are indeterminate.

\subsection{Results}

\begin{figure}[tp]
\centering
\includegraphics[width=0.8\columnwidth]{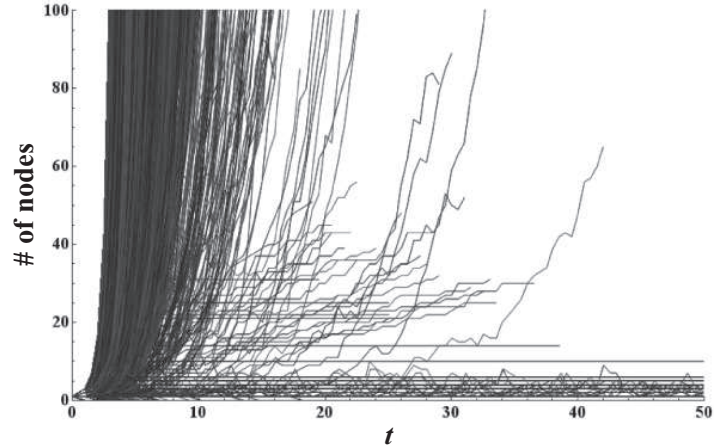}
\caption{Growth curves of randomly selected 10\% of the 50000
  independent simulation runs (5 runs $\times$ 10000 rules).}
\label{growthcurves}
\end{figure}

We first studied the growth characteristics of GNA and their
differences between different rules. Figure \ref{growthcurves}
presents sample growth curves superposed in a single plot, showing
temporal changes in number of nodes over simulated time. Several
distinct types of growth patterns are already visible in this
plot. Curves that go nearly flat along the $t$ axis indicate that the
GNA for these cases did not grow at all. Many other rules showed rapid
exponential growth processes (dense bundle of sharply rising curves on
the left). Between these two, there are relatively fewer intermediate
cases that exhibit either slow, fluctuating growth, or even linear
growth, which are qualitatively different from other growth curves.

We extracted the growth characteristics of each rule from its time
series data by fitting to them a hypothetical growth model $N_t \sim
(t+1)^k e^{rt}$ using the least squares method, where $k$ and $r$ are
the estimated order of polynomial growth and the estimated rate of
exponential growth, respectively. For each rule, these values were
calculated with five independent simulation runs and then their
averages were used for the analysis. We excluded rules in the form of
``***0'' or ``0**2'' (where `*' can be any single-digit number) that
caused immediate node extinction and hence failure of nonlinear
fitting. This filtering excluded 1100 rules, leaving a total of 8900
(out of 10000) rules that were used in the following plots.

\begin{figure}[tp]
\centering
\includegraphics[width=\columnwidth]{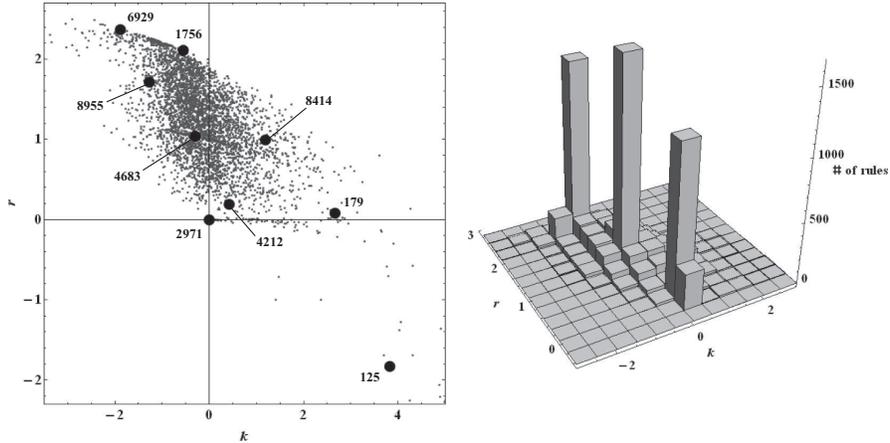}
\caption{Left: Distribution of growth characteristics, estimated order
  of polynomial growth $k$ and estimated rate of exponential growth
  $r$, of the 8900 GNA rules (excluding rules in the form of ``***0''
  or ``0**2'' that caused immediate node extinction and hence failure
  of nonlinear fitting). Sample cases shown in Fig.~\ref{shapes} are
  indicated with large black dots, accompanied with the corresponding
  rule numbers. Right: 3-D histogram of growth characteristics in the
  parameter area $-3<k<3$ and $0<r<3$. It is clearly seen that there
  are three distinct peaks, which correspond to non-growers,
  exponential growers by binary node divisions, and exponential
  growers by tertiary node divisions.}
\label{krplot}
\end{figure}

Figure \ref{krplot} (left) shows the distribution of the growth
characteristics ($k$ and $r$) of the 8900 GNA rules. The distribution
is continuously spread mostly in the first and second quadrants, in
which there are a couple of visually identifiable dense clusters. The
slightly slanted linear cluster near the top of the second quadrant
corresponds to rules that make GNA grow exponentially through
continuous tertiary node divisions. The other slanted linear cluster
located around $(k,r)=(0,1)$ corresponds to rules that make GNA grow
exponentially through continuous binary divisions. Between and around
these two clusters there are many other rules that show intermediate
exponential growth rates. A relatively thin linear cluster at $k>0$
and $r \sim 0$ is considered of non-growing or polynomially growing
GNA rules. Most of the GNA rules belong to one of these three
clusters, as seen in the histogram on the right. Finally, the sparse
distribution of rules in the fourth quadrant are the ones that lead to
node extinction.

\begin{sidewaysfigure}[p]
\centering
\includegraphics[width=\columnwidth]{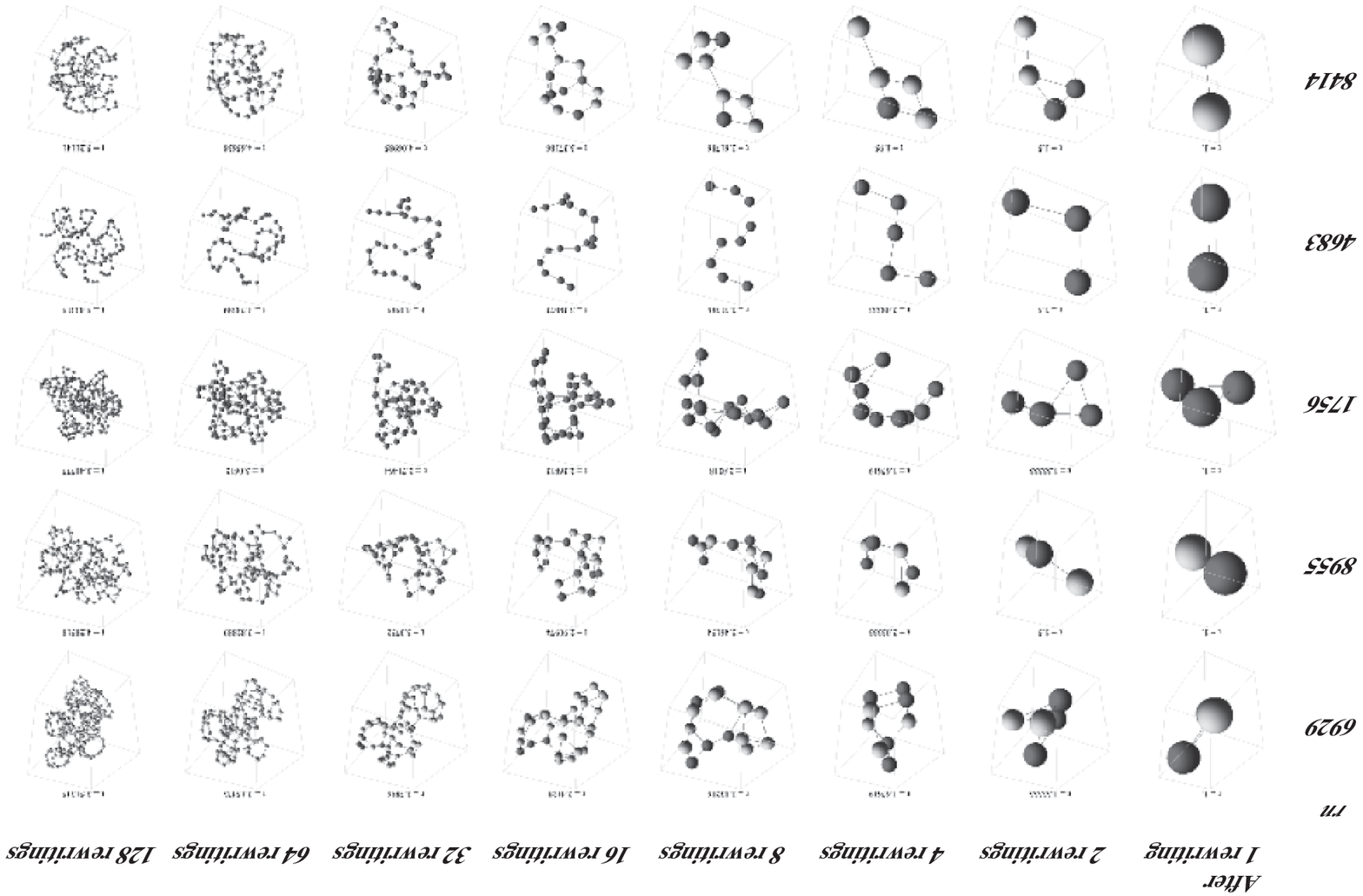}
\end{sidewaysfigure}

\begin{sidewaysfigure}[p]
\centering
\includegraphics[width=\columnwidth]{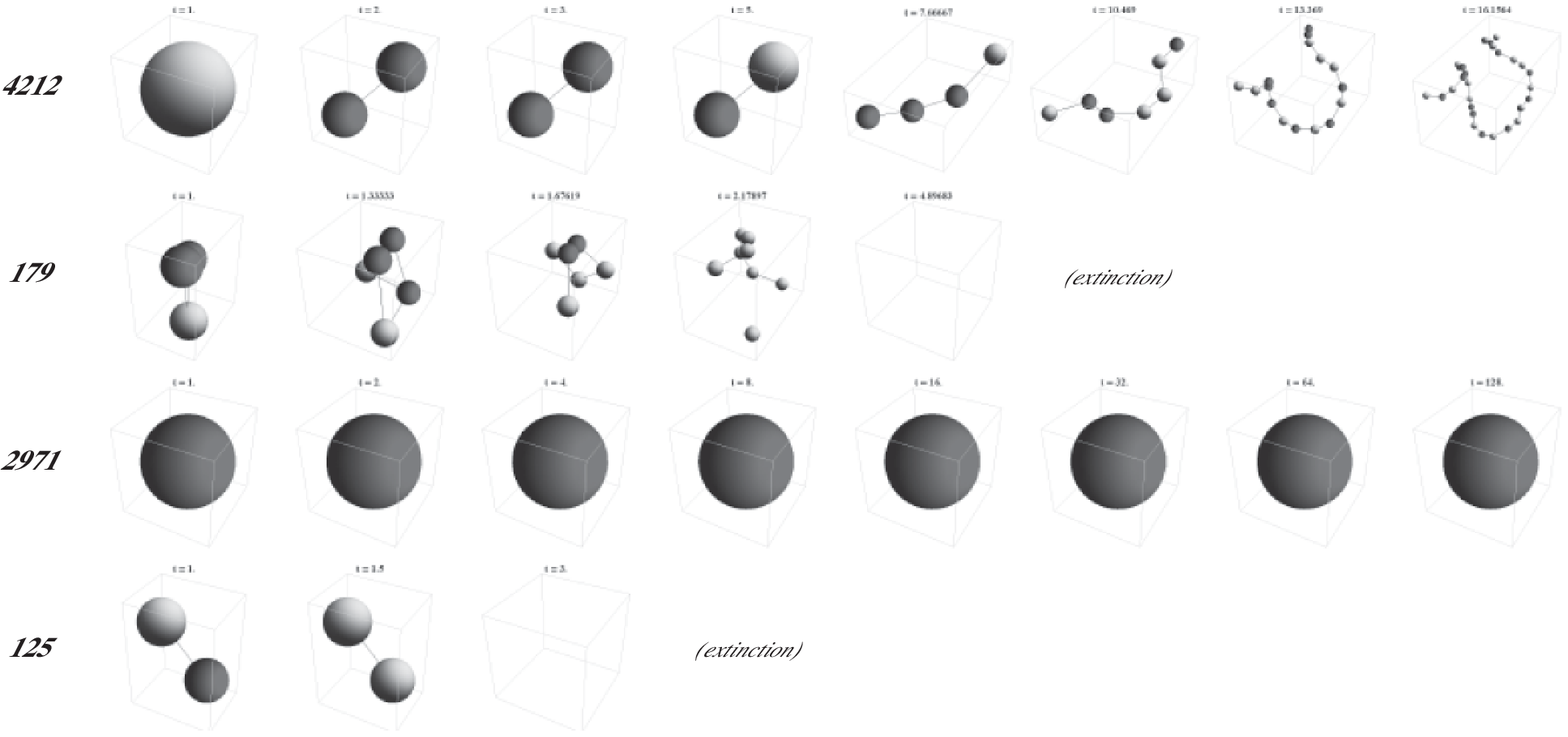}
\caption{Sample growth patterns of several GNA rules. Each row
  presents one specific simulation run for a particular GNA rule
  (indicated on the left by $rn$). Each image shows an actual GNA
  configuration after specific times of rewriting events. Note that
  the number of rewritings does not necessarily scale along the
  simulated time, which is given at the top of each image. Dark gray
  dots represent nodes in state 0, while light gray ones in state 1.}
\label{shapes}
\end{sidewaysfigure}

Figure \ref{shapes} shows actual growth patterns of several rule
samples (indicated by large black dots in Fig.~\ref{krplot}), which
confirms topological diversity generated even within this restricted
set of binary-state GNA rules. The first five rows ($rn=$6929, 8955,
1756, 4683 and 8414) are the samples of exponentially growing
rules. For $rn=$1756 and 4683, every rewriting event exclusively
causes tertiary and binary node divisions and forms planar and linear
structures, respectively, where node states remain homogeneous and do
not change at all. On the other hand, for $rn=$6929, 8955 and 8414,
state-1 nodes appear at the beginning of simulation and the node
states influence the network growth processes. Such interaction
between node states and network topology results in a final GNA
configuration with non-homogeneous node state distribution and a
growth rate that is different from those of homogeneous network
growth. The rest are the examples that do not show exponential growth,
among which $rn=$4212 uniquely demonstrates a very slow growth of a
linear structure driven by a complicated interaction between state-0
and state-1 nodes on it.

\begin{figure}[tp]
\centering
\includegraphics[width=\columnwidth]{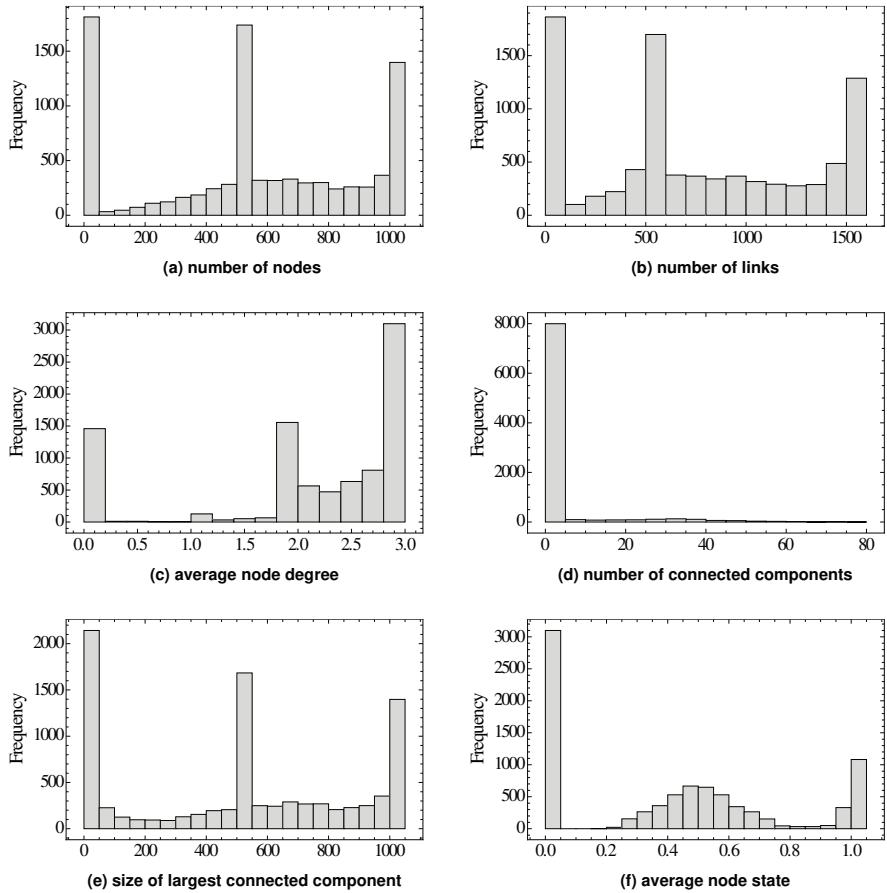}
\caption{Histograms of rule frequencies over six topological
  characteristics of the final GNA configurations. Each characteristic
  was calculated by averaging measurements obtained from five
  independent simulation runs for each rule. 8617 GNA rules after
  filtering (see text) were used to produce these plots.}
\label{histograms}
\end{figure}

We also investigated the topological characteristics of the final GNA
configurations obtained at the end of each simulation run. For this
purpose, we additionally excluded rules that always ended up with node
extinction, because average node degrees or states cannot be defined
for such rules. As a result, we used 8617 rules for the following
analyses. Figure \ref{histograms} shows the histograms of rule
frequencies arranged in terms of six topological characteristics
(described earlier) of the final GNA configuration. Three distinct
peaks are commonly seen in (a), (b), (c) and (e) of these plots. These
three peaks correspond to two types of exponential growers (by
tertiary and binary node divisions) and non-growers. Between these
peaks other cases distribute with relatively lower frequencies. Plot
(d) indicates that most rules produce connected network structures
only. In terms of the node state distribution, plot (f) shows that
many GNA rules produce networks which are homogeneous regarding node
states (represented by two peaks at 0.0 and 1.0) but other rules
produce heterogeneous state distributions as well (represented by a
gentle peak around 0.5). The distribution in (f) is asymmetric because
we used a single node of state 0 as an initial condition for all the
simulations.

\begin{figure}[tp]
\centering
\includegraphics[width=\columnwidth]{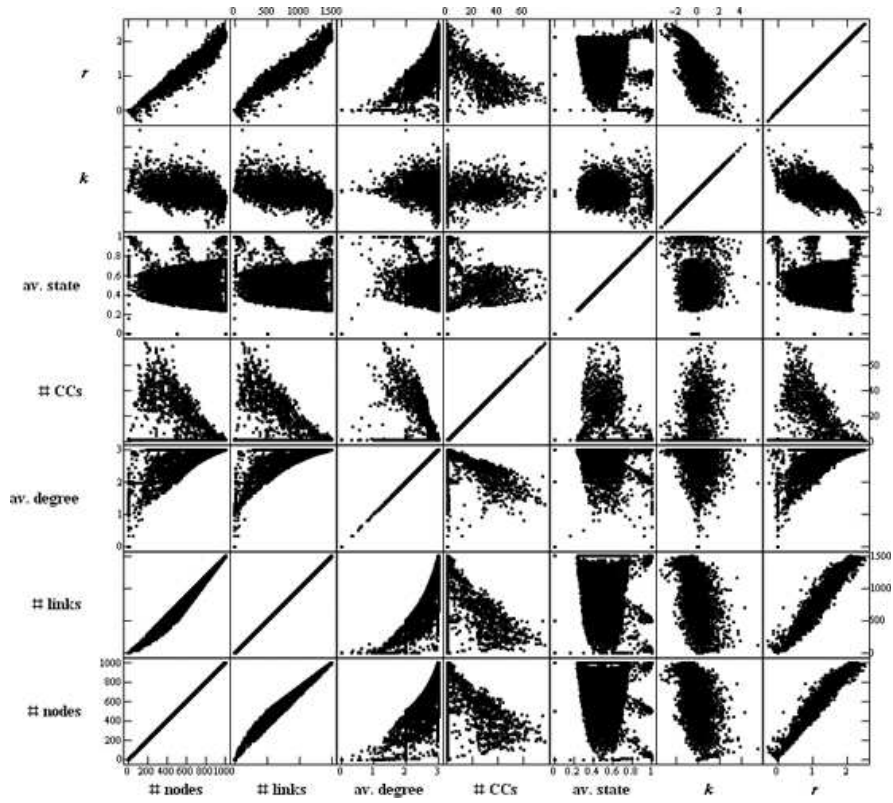}
\caption{Scatter plot matrix showing $7 \times 7 = 49$ scatter plots,
  each of which visually shows correlation between two of the
  following seven characteristics of GNA. From left (bottom) to right
  (top): number of nodes, number of links, average node degree, number
  of connected components, average node state, estimated order of
  polynomial growth $k$, and estimated rate of exponential growth
  $r$.}
\label{scatterplots}
\end{figure}

Figure \ref{scatterplots} is a scatter plot matrix made of $7 \times 7
= 49$ scatter plots, each of which visually shows correlation between
two of the seven characteristics described above: number of nodes,
number of links, average node degree, number of connected components,
average node state, estimated order of polynomial growth $k$, and
estimated rate of exponential growth $r$. The size of largest
connected components was not included because it is strongly
correlated with the number of nodes as most of the networks were well
connected (see Fig.~\ref{histograms} (d), as well as (a) and
(e)). This matrix gives several interesting observations. There is a
simple correlation between number of nodes, number of links and
average node degree for obvious reasons, as already reported in our
previous work \cite{sayama07}. More importantly, average node states
have significant impacts on other properties of GNA, as seen in the
fifth column/row of the matrix. For networks whose node states are
homogeneous (i.e., average node state $\sim$ 0 or 1), there is always
only one local situation possible: a node of state 0 (or 1) surrounded
by nodes of the same state. For such a network to remain in
homogeneous states while staying away from node extinction, there are
only three possible outcomes (tertiary division with state preserved,
binary division with state preserved, or absolutely no change). This
necessarily results in only three values possible for number of nodes,
number of links and estimated rate of exponential growth, for average
node state $\sim$ 0 or 1. It is also notable that the largest numbers
of connected components are achieved when the average node states take
intermediate values. This suggests that node states play a critical
role in determining when and where a node should disappear to cut the
network and increase the number of connected components. Without such
state-driven control of node disappearance, the nodes would easily
become extinct.

Finally, we conducted principal component analysis (PCA) on the
distribution of results in a seven-dimensional vector space created by
the seven characteristics used in Fig.~\ref{scatterplots}. Data were
rescaled before the analysis so that the standard deviation was one in
each dimension. As a result, we extracted four important dimensions in
the data distribution (Table \ref{PCAresults}). The primary dimension
is strongly correlated with number of nodes, number of links, average
node degree, and estimated rate of exponential growth $r$, which may
be understood as a factor relevant to general topological growth. The
secondary dimension is strongly correlated to number of connected
components, average node state, and estimated order of polynomial
growth $k$, which may be understood as a factor related to node
disappearances caused by state changes. Note that the basis vector of
this dimension happened to be taken in opposite direction to its
correlated characteristics, so the lower value in this dimension means
greater number of connected components, higher average node state, and
higher order of polynomial growth.

\begin{table}[tp]
\centering
\caption{Results of principal component analysis (PCA) applied to the
  same data shown in Fig.~\ref{scatterplots}. Components and
  eigenvalues in bold face indicate four important dimensions.}
\label{PCAresults}
\begin{tabular}{ccrrrrrrr}
\hline
\hline
{\bf Component} & {\bf Eigenvalue} & \multicolumn{7}{c}{\bf Eigenvector} \\
& &
\multicolumn{1}{c}{\# of} &
\multicolumn{1}{c}{\# of} &
\multicolumn{1}{c}{Av.\ node} &
\multicolumn{1}{c}{\# of} &
\multicolumn{1}{c}{Av.\ node} &
\multicolumn{1}{c}{$k$} &
\multicolumn{1}{c}{$r$} \\
 & & 
\multicolumn{1}{c}{nodes} &
\multicolumn{1}{c}{links} &
\multicolumn{1}{c}{degree} &
\multicolumn{1}{c}{CCs} &
\multicolumn{1}{c}{state} & & \\
\hline
{\bf 1} & {\bf 4.014} &  0.490 &  0.485 &  0.441 & -0.087 &  0.082 & -0.265 &  0.495 \\
{\bf 2} & {\bf 1.203} & -0.021 &  0.005 & -0.269 & -0.642 & -0.603 & -0.388 &  0.034 \\
{\bf 3} & {\bf 0.895} &  0.122 &  0.108 &  0.028 &  0.584 & -0.763 &  0.204 &  0.085 \\
{\bf 4} & {\bf 0.718} &  0.105 &  0.121 &  0.192 & -0.484 & -0.109 &  0.831 & -0.018 \\
     5  &      0.151  & -0.234 & -0.383 &  0.828 & -0.071 & -0.186 & -0.177 & -0.207 \\
     6  &      0.019  &  0.490 & -0.768 & -0.101 & -0.018 &  0.015 &  0.073 &  0.392 \\
     7  &      0.001  & -0.662 & -0.034 &  0.004 & -0.002 & -0.002 &  0.102 &  0.741 \\
\hline
\end{tabular}
\end{table}

\begin{figure}[tp]
\centering
\includegraphics[width=0.7\columnwidth]{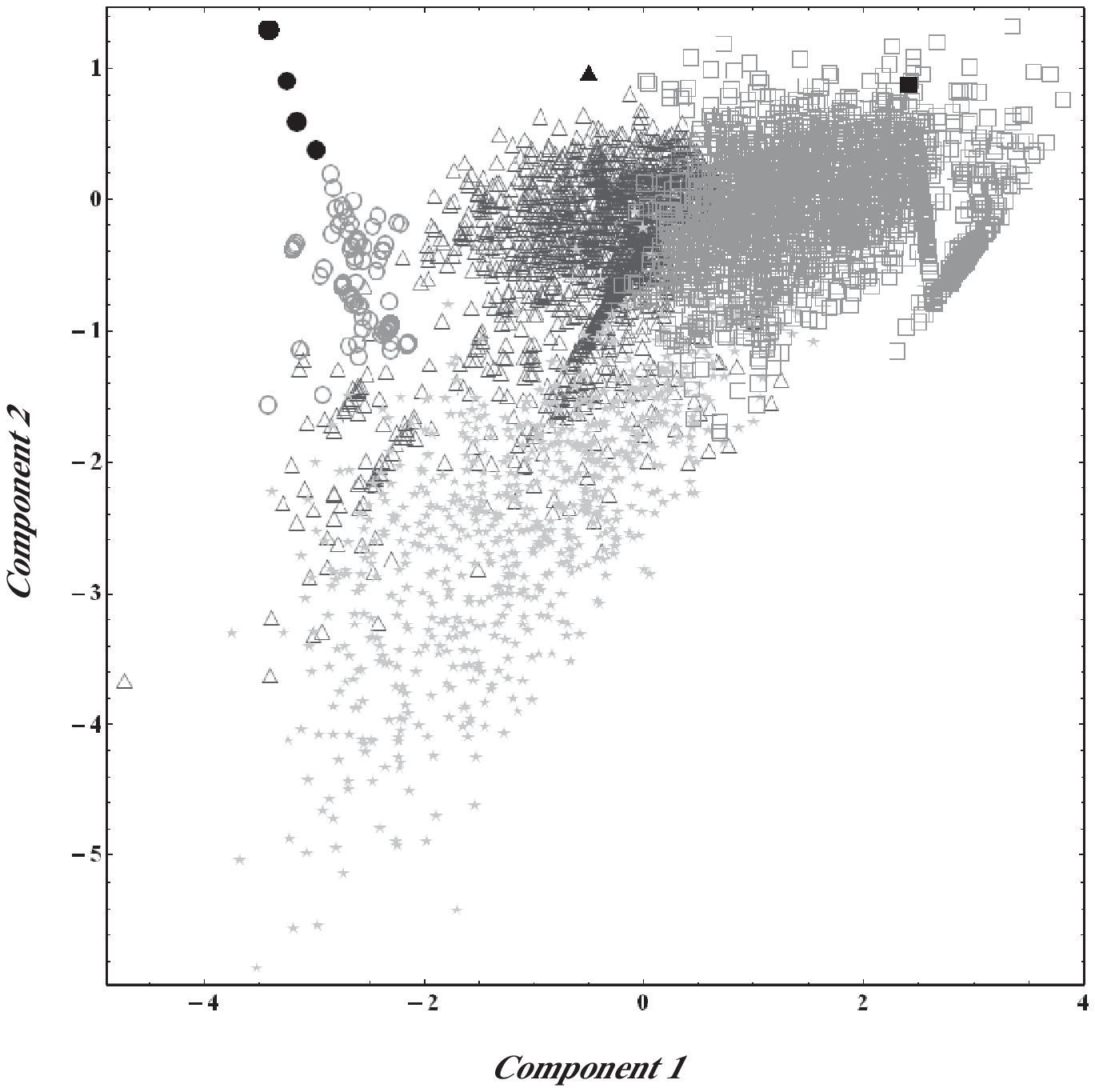}\\~\\
\includegraphics[width=0.7\columnwidth]{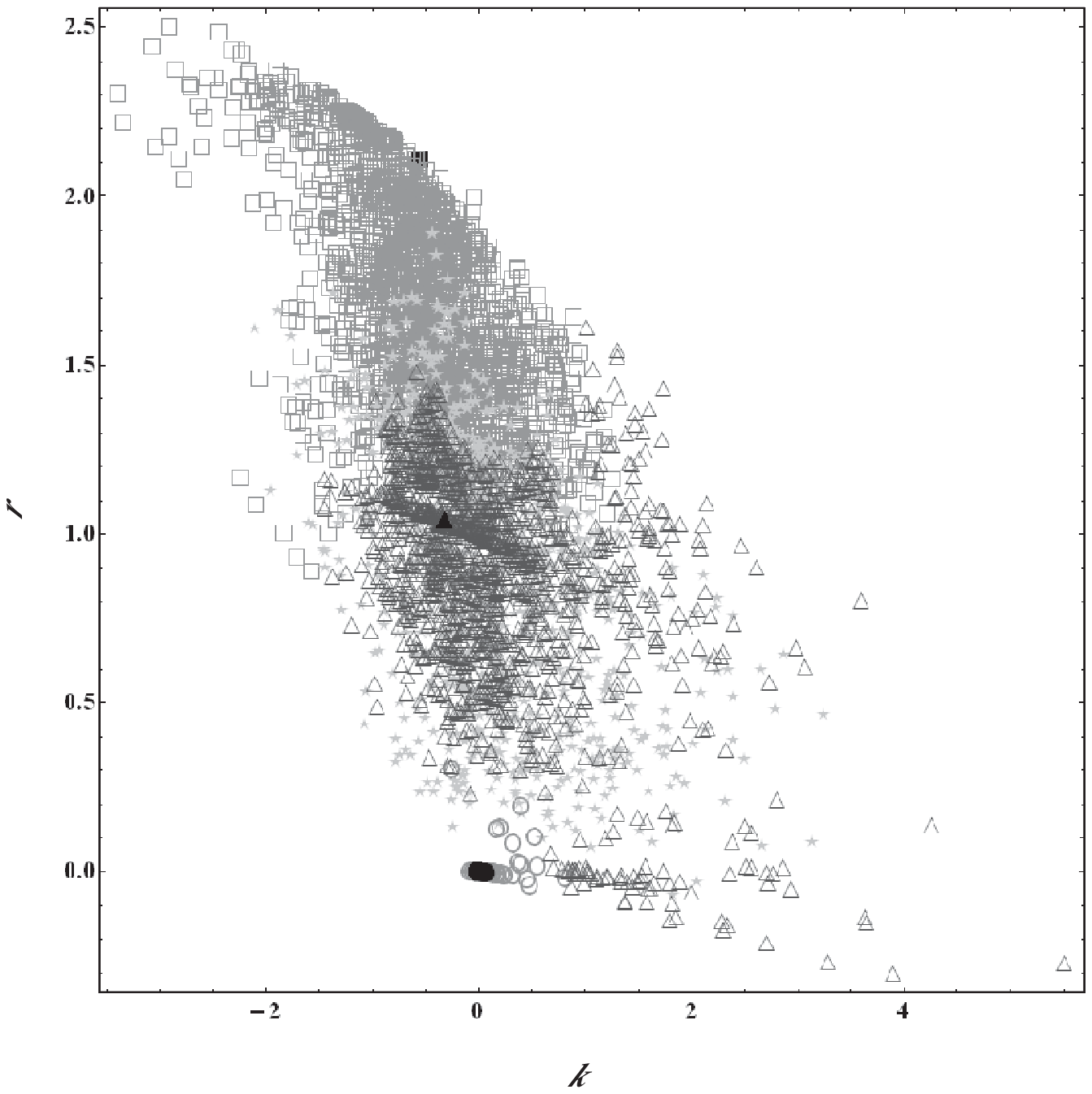}
\caption{Results of hierarchical clustering of the data distribution
  conducted in a dimension-reduced vector space. Top: Clusters
  projected to a two-dimensional space using the primary and secondary
  dimensions detected by PCA. Bottom: Same results mapped in the
  $k$-$r$ space in the same way as in Fig.~\ref{krplot}. Numbers of
  rules in these clusters are as follows: Filled circle 1103, filled
  triangle 1000, filled square 1000, open circle 412, open triangle
  1812, open square 2529, and star 761.}
\label{clusters}
\end{figure}

We further applied Ward's minimum-variance hierarchical clustering
algorithm to the data distribution in a vector space whose dimensions
were reduced from seven to four according to the results of PCA. The
clustering results were split into seven clusters as shown in
Fig.~\ref{clusters}, where the top plot presents the results in a
two-dimensional space using the primary and secondary dimensions
detected by PCA, whereas the bottom plot maps the same results in the
$k$-$r$ space in the same way as in Fig.~\ref{krplot}.

Rules in each cluster were manually sampled and inspected in further
detail to see what kind of common dynamics exist within each type,
which revealed the following: The first three clusters, filled circles
(1103 rules), filled triangles (1000 rules) and filled squares (1000
rules), share exactly the same growth characteristics within each
cluster so that they appear as a point in the $k$-$r$ plot
(Fig.~\ref{clusters}, bottom). Specifically, the filled circles are
non-growers without state changes or with regular state alterations
between 0 and 1 (e.g., $rn=$2971), while the filled triangles and the
filled squares are exponential growers without state changes, growing
solely by binary (e.g., $rn=$4683) and tertiary (e.g., $rn=$1756) node
divisions, respectively. The other three clusters denoted by open
markers involve active node state changes that influence their growth
patterns. Specifically, the open circles (412 rules) show very slow or
even no growth (e.g., $rn=$4212), while the open triangles (1812
rules) and the open squares (2529 rules) show exponential-like growth
predominantly by binary (e.g., $rn=$8414) and tertiary (e.g.,
$rn=$6929 and 8955) node divisions, respectively. Finally, the cluster
denoted by light gray stars (761 rules) involve active node state
changes and frequent node disappearances, typically producing more
than one connected components (e.g., $rn=$179).

These results altogether demonstrate the diversities of potential
dynamics of simple GNAs, in both topology and temporal evolution. Of
particular importance compared to other network growth models is the
possibility of interaction between network topology and node state
distribution, which is key to nontrivial dynamics observed in the
types that involve active node state changes.

\section{Conclusion}

We proposed Generative Network Automata as a new generalized framework
for the modeling of complex dynamical networks, with which one can
uniformly describe both state transitions and autonomous topological
transformations using repetitive graph rewritings. We explored
possible dynamics of simple binary-state GNA and observed several
distinct types of topologies and growth patterns that emerged from
local rewriting rules, where dynamic state changes were coupled with
topological changes in some types.

The work presented here had a couple of limitations that must be
noted. One was that we employed several restrictions on possible rule
sets to keep the search space small. For example, we assumed that the
extraction mechanism $E$ randomly picks a node from the network, which
avoided the computationally inefficient subgraph-isomorphism problem
that would need to be solved for other types of extraction mechanisms
that look for particular topological features. The other limitation
was the small network size. We experimented with GNAs whose size was
up to 1000 nodes, which is significantly smaller than many real-world
complex networks being investigated today. We realize that, to enable
unrestricted GNA modeling and simulation at a significantly larger
scale, several technical issues will need to be addressed in a
computationally efficient way, including:
\begin{enumerate}
\item How to represent and rewrite large GNA configurations
\item How to extract subGNAs that match given patterns from a large
  GNA configuration
\item How to keep track of statistical/dynamical properties of GNAs
  during simulation with minimum computational overheads
\item How to embed complex GNAs in a 2-D or 3-D visualization space in
  a visually meaningful manner
\item How to derive the optimal rule set that best explains the
  network evolution given by experimental data
\end{enumerate}
Some of these problems apparently involve intractable computational
complexity if exact solutions are sought\footnote{The
  subgraph-isomorphism problem is known to be NP-complete.}. We are
therefore working to develop computationally practical solutions to
these problems by using appropriate approximations and heuristics.

We hope that GNA will help formulate many distinct complex systems in
the same ``format'', enabling one to compare those systems
systematically, to identify their commonness and uniqueness, and to
actively exchange knowledge between different fields beyond
disciplinary boundaries. We anticipate several areas of immediate
applications, including (a) ecology and epidemiology modeling where
organisms and pathogens actively reshape their habitat structure
(e.g., niche construction, effects of host survivability in
epidemiological networks), (b) social network modeling where
individual states and behaviors modify the network topology (e.g.,
evolution of social ties, self-organization of collective knowledge
among people), and (c) biologically inspired engineering design where
local rewriting rules can be exploited as a means to indirectly
control the emergent dynamics of artifacts that develop and
self-organize over time.

\end{document}